\documentclass{elsart}
\usepackage{epsfig}
\usepackage{color}
\usepackage{amssymb}
\newcommand{\aff}[2]{Dipartimento di Fisica dell'Universit\`a #1 e Sezione INFN, #2, Italy.}
\newcommand{\affd}[1]{Dipartimento di Fisica dell'Universit\`a e Sezione INFN, #1, Italy.}
\begin{document}
\begin{frontmatter}
\date{}

\title{Study of the radiative decay $\phi\to a_0$(980)$\gamma$ with the KLOE
  detector}

\collab{The KLOE Collaboration}

\author[Na]{F.~Ambrosino},
\author[Frascati]{A.~Antonelli},
\author[Frascati]{M.~Antonelli},
\author[Frascati]{F.~Archilli},
\author[Roma3]{C.~Bacci},
\author[Karlsruhe]{P.~Beltrame},
\author[Frascati]{G.~Bencivenni},
\author[Frascati]{S.~Bertolucci},
\author[Roma1]{C.~Bini},
\author[Frascati]{C.~Bloise},
\author[Roma3]{S.~Bocchetta},
\author[Roma1]{V.~Bocci},
\author[Frascati]{F.~Bossi},
\author[Roma3]{P.~Branchini},
\author[Roma1]{R.~Caloi},
\author[Frascati]{P.~Campana},
\author[Frascati]{G.~Capon},
\author[Na]{T.~Capussela},
\author[Roma3]{F.~Ceradini},
\author[Frascati]{S.~Chi},
\author[Na]{G.~Chiefari},
\author[Frascati]{P.~Ciambrone},
\author[Frascati]{E.~De~Lucia},
\author[Roma1]{A.~De~Santis},
\author[Frascati]{P.~De~Simone},
\author[Roma1]{G.~De~Zorzi},
\author[Karlsruhe]{A.~Denig},
\author[Roma1]{A.~Di~Domenico},
\author[Na]{C.~Di~Donato},
\author[Pisa]{S.~Di~Falco},
\author[Roma3]{B.~Di~Micco},
\author[Na]{A.~Doria},
\author[Frascati]{M.~Dreucci},
\author[Frascati]{G.~Felici},
\author[Frascati]{A.~Ferrari},
\author[Frascati]{M.~L.~Ferrer},
\author[Frascati]{G.~Finocchiaro},
\author[Roma1]{S.~Fiore},
\author[Frascati]{C.~Forti},
\author[Roma1]{P.~Franzini},
\author[Frascati]{C.~Gatti},
\author[Roma1]{P.~Gauzzi},
\author[Frascati]{S.~Giovannella},
\author[Lecce]{E.~Gorini},
\author[Roma3]{E.~Graziani},
\author[Pisa]{M.~Incagli},
\author[Karlsruhe]{W.~Kluge},
\author[Moscow]{V.~Kulikov},
\author[Roma1]{F.~Lacava},
\author[Frascati]{G.~Lanfranchi},
\author[Frascati,StonyBrook]{J.~Lee-Franzini},
\author[Karlsruhe]{D.~Leone},
\author[Frascati]{M.~Martini},
\author[Na]{P.~Massarotti},
\author[Frascati]{W.~Mei},
\author[Na]{S.~Meola},
\author[Frascati]{S.~Miscetti},
\author[Frascati]{M.~Moulson},
\author[Frascati]{S.~M\"uller},
\author[Frascati]{F.~Murtas},
\author[Na]{M.~Napolitano},
\author[Roma3]{F.~Nguyen},
\author[Frascati]{M.~Palutan},
\author[Roma1]{E.~Pasqualucci},
\author[Roma3]{A.~Passeri},
\author[Frascati,Energ]{V.~Patera},
\author[Na]{F.~Perfetto},
\author[Lecce]{M.~Primavera},
\author[Frascati]{P.~Santangelo},
\author[Na]{G.~Saracino},
\author[Frascati]{B.~Sciascia},
\author[Frascati,Energ]{A.~Sciubba},
\author[Pisa]{F.~Scuri},
\author[Frascati]{I.~Sfiligoi},
\author[Frascati]{T.~Spadaro},
\author[Roma1]{M.~Testa},
\author[Roma3]{L.~Tortora},
\author[Roma1]{P.~Valente},
\author[Karlsruhe]{B.~Valeriani},
\author[Frascati]{G.~Venanzoni},
\author[Frascati]{R.Versaci},
\author[Frascati,Beijing]{G.~Xu}

\address[Frascati]{Laboratori Nazionali di Frascati dell'INFN, 
Frascati, Italy.}
\address[Karlsruhe]{Institut f\"ur Experimentelle Kernphysik, 
Universit\"at Karlsruhe, Germany.}
\address[Lecce]{\affd{Lecce}}
\address[Na]{Dipartimento di Scienze Fisiche dell'Universit\`a 
``Federico II'' e Sezione INFN,
Napoli, Italy}
\address[Pisa]{\affd{Pisa}}
\address[Energ]{Dipartimento di Energetica dell'Universit\`a 
``La Sapienza'', Roma, Italy.}
\address[Roma1]{\aff{``La Sapienza''}{Roma}}
\address[Roma3]{\aff{``Roma Tre''}{Roma}}
\address[StonyBrook]{Physics Department, State University of New 
York at Stony Brook, USA.}
\address[Beijing]{Permanent address: Institute of High Energy 
Physics of Academica Sinica,  Beijing, China.}
\address[Moscow]{Permanent address: Institute for Theoretical 
and Experimental Physics, Moscow, Russia.}

\begin{abstract}
A sample of $1.25\times 10^9$ $\phi$ decays, collected with the KLOE
detector at the Frascati $\phi$-factory DA$\Phi$NE at center of mass
energy $\sim M_{\phi}$, has been used to study the radiative decay
$\phi\to\eta\pi^0\gamma$.
This decay is dominated by $\phi\to a_0(980)\gamma$. 
Two decay chains, corresponding to $\eta\to\gamma\gamma$ and
$\eta\to\pi^+\pi^-\pi^0$, have been selected.
We found respectively: $Br(\phi\to\eta\pi^0\gamma)=(6.92\pm 0.10_{stat.} \pm
0.20_{syst.})\times 10^{-5}$ and $(7.19\pm 0.17_{stat.}\pm0.24_{syst.})\times 10^{-5}$.
The $\eta\pi^0$ invariant mass distributions have been fitted to obtain the
relevant $a_0$(980) parameters. 
\end{abstract}

\begin{keyword}
$e^+e^-$ collisions, $\phi$ radiative decays, Scalar mesons
\end{keyword} 
\end{frontmatter}

\maketitle
\section{Introduction}
The scalar isovector meson $a_0$(980), as well as the isoscalar $f_0$(980), 
is not easily interpreted as an ordinary q\=q meson belonging to the $^3P_0$
nonet. 
Alternative hypotheses have been proposed: q\=qq\=q states\cite{jaffe},
K\=K bound states\cite{isgur}, or dynamically generated
resonances\cite{oset}. \\
The radiative decay $\phi\to\eta\pi^0\gamma$ is dominated by the exchange
of the $a_0$(980) in the intermediate state ($\phi\to a_0\gamma$ with 
$a_0\to\eta\pi^0$); the contribution of the interfering process 
$\phi\to\rho^0\pi^0$ is small due to the small $\rho^0\to\eta\gamma$ coupling.
According to the possible different structures, the branching ratio of
$\phi\to a_0\gamma$ can vary from $10^{-5}$ in the q\=q or K\=K case
to $10^{-4}$ in the q\=qq\=q hypothesis.
The mass shape is also expected to depend on the meson structure.
The process has been observed and measured by SND\cite{snd} and
CMD-2\cite{cmd2} at VEPP-2M and by KLOE\cite{kloe} during its first period
of data taking with limited statistics. \\
 In this paper we present preliminary results of the analysis of the decay
$\phi\to\eta\pi^0\gamma$ performed with the KLOE detector,
operated at the Frascati $\phi$-factory DA$\Phi$NE\cite{daphne}, using a
sample of 414 pb$^{-1}$ of the 2001-2002 data taking, corresponding to
$1.25\times 10^9$ $\phi$ produced, about a factor 20 more in statistics
with respect to the previous result\cite{kloe}.
The decay chains corresponding to $\eta\to\gamma\gamma$ and
$\eta\to\pi^+\pi^-\pi^0$ have been selected to extract the branching ratio;
the former is characterized by a higher branching ratio and large
 background, the latter has lower branching ratio but a smaller 
background contamination. \\
The two invariant mass spectra have then been used to evaluate the
relevant $a_0$ parameters (mass and couplings) through a combined fit. 

\section{The KLOE detector}
The KLOE detector consists of a large cylindrical drift chamber\cite{K-dc},
surrounded by a lead/scintillating-fiber sampling calorimeter\cite{K-emc},
both immersed in a solenoi\-dal magnetic field of 0.52 T with the axis
parallel to the beams. 
Two small calorimeters\cite{K-qcal} are wrapped around the quadrupoles
of the low-$\beta$ insertion to complete the detector hermeticity. \\
The drift chamber provides a measurement of charged tracks momentum with a
resolution of $\delta p_\perp/p_\perp$=0.4\%, and of decay 
vertices with an accuracy of 3 mm. \\
The calorimeter covers 98$\%$ of the total
solid angle and provides measurements of energy, time and position of
photons with accuracies of $\sigma_{E}/E = 0.057/{\sqrt{E \ ({\rm GeV})}}$
and $\sigma_{t} = 57 \ {\rm ps} /{ \sqrt{E \ ({\rm GeV})}} \oplus 50 \
{\rm ps}$, respectively.
A photon is defined as a cluster of energy deposits in the calorimeter 
not associated to a charged particle. \\
The trigger\cite{K-trigger} uses information from both the
calorimeter and the drift chamber. The calorimeter trigger requires two
local energy deposits.

\section{$\phi\to\eta\pi^0\gamma$ with $\eta\to\gamma\gamma$}
\begin{figure}[htb]
\begin{tabular}{cc}
\hskip -.7cm
\mbox{\epsfig{file=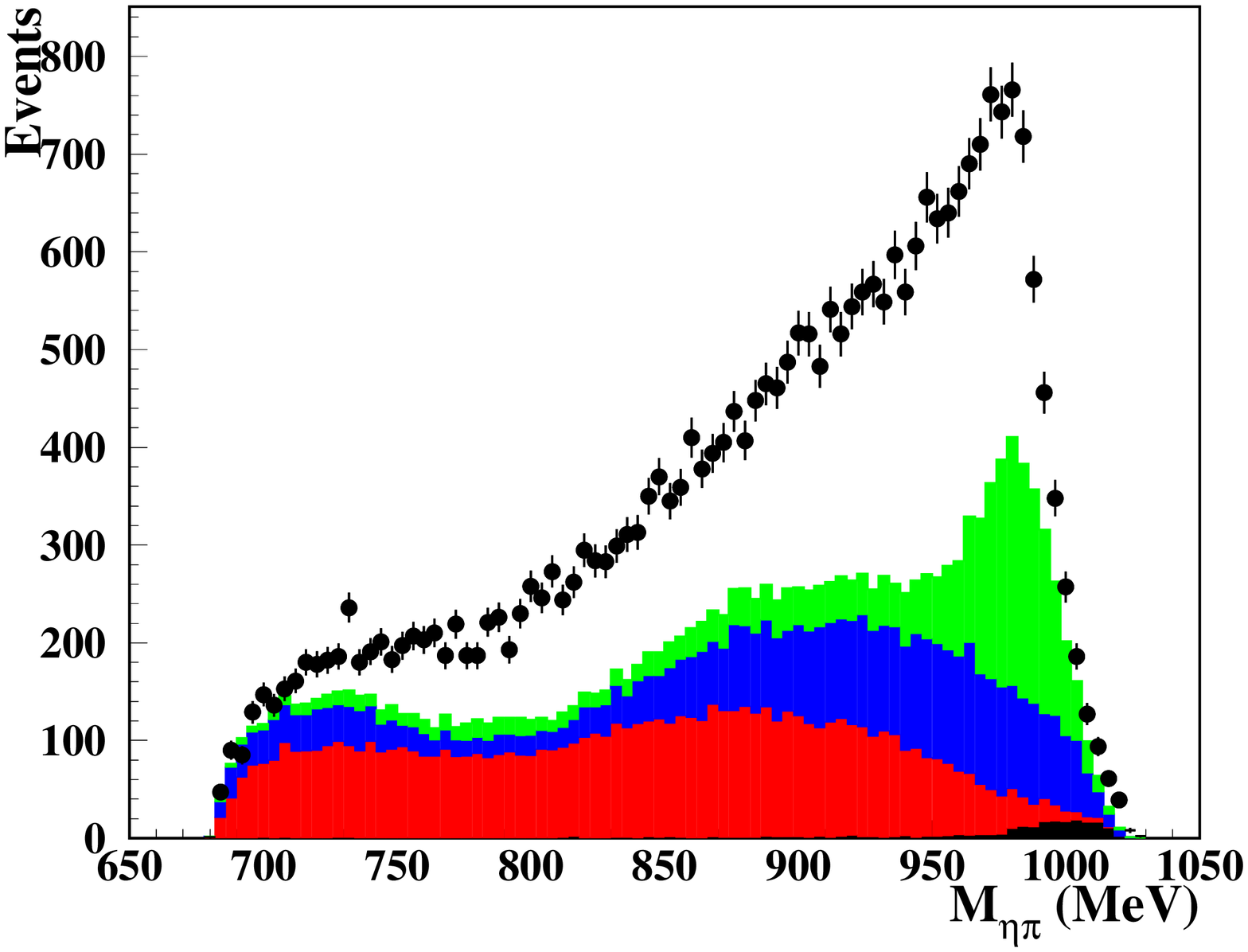, width=7cm}} &
\mbox{\epsfig{file=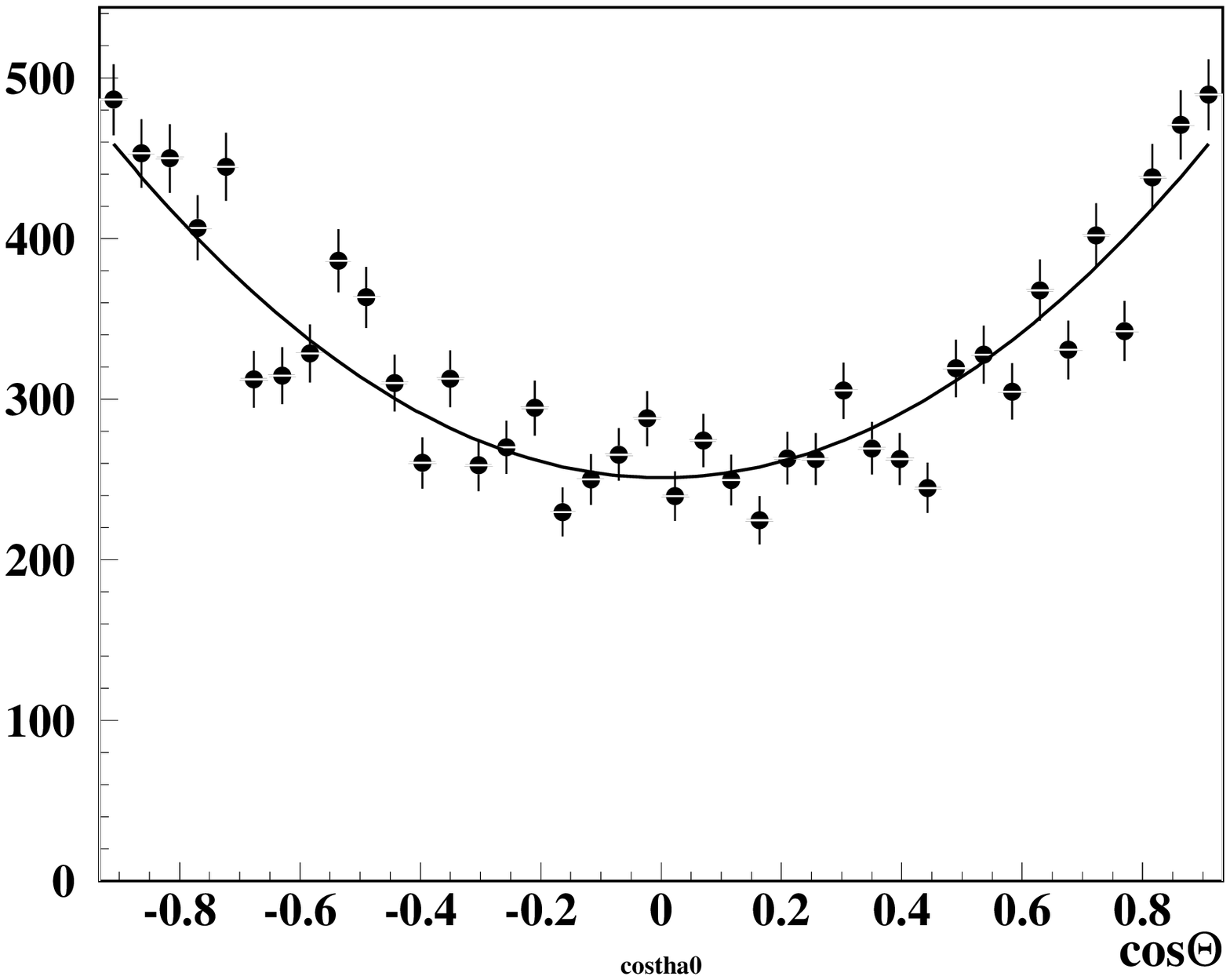, width=6.4cm}} \\
\end{tabular}
\caption{Left: $M_{\eta\pi^0}$ distribution before background
    subtraction; dots are data, and coloured histograms are the background
    contributions evaluated by MC (black: $\phi\to\eta\gamma$ with
    $\eta\to\gamma\gamma$, blue: $\phi\to\eta\gamma$ with $\eta\to
    3\pi^0$, green: $e^+e^-\to\omega\pi^0$, red:$\phi\to f_0\gamma$). 
    Right: polar angle distribution of the recoil photon after the
    background subtraction.} 
\label{neutri}
\end{figure}

Events from this decay chain are characterized by five prompt photons
in the final state, without any charged track in the drift chamber.
A prompt photon is defined as a calorimeter cluster satisfying the
condition $|t-r/c| < 5\sigma_t(E)$, where $r$ is the distance of the impact
point on the calorimeter from the beam interaction point, $t$ is the
arrival time, $c$ is the speed of light. \\
The main background processes are:
\begin{enumerate}
  \item \label{f0} $\phi\to\pi^0\pi^0\gamma$, dominated by $\phi\to f_0\gamma$
  \item \label{wpn} $e^+e^-\to\omega\pi^0$ with $\omega\to\pi^0\gamma$
  \item \label{eg7} $\phi\to\eta\gamma$ with $\eta\to\pi^0\pi^0\pi^0$ 
  \item \label{eg3} $\phi\to\eta\gamma$ with $\eta\to\gamma\gamma$
\end{enumerate}
Events from process (\ref{eg7}) can be wrongly reconstructed as five photon
ones due to cluster merging or loss, while events from process \ref{eg3} can
mimic the signal due to photon splittings or to accidental 
clusters in the calorimeter. 
The background reduction proceeds through the following steps: {\it (i)}
a kinematic fit on the events is performed, with the constraints of the
4-momentum conservation and $|t-r/c|=0$ for each prompt photon; {\it (ii)}
the best photon pairing to $\pi^0$'s and $\eta$'s is searched for in the
signal and background hypotheses; and then {\it (iii)} a second kinematic
fit is applied by imposing also the constraints of the masses of the
intermediate particles.
The $\chi^2$ of these fits and other kinematical variables are used
to cut the background events. 
The overall selection efficiency is 39\%.
The background has been evaluated on control samples dominated
by the different background processes.
By comparing the data with the Monte Carlo (MC) simulation of the
experiment, the amount of each background, to be subtracted from the final
sample, has been obtained(see fig.\ref{neutri}, left). \\
The selected sample consists of 29601 events; the total background amounts to
$16332\pm 86$ events, from which we extract the signal events to be $13269\pm 192$.
The Br is obtained by normalizing the signal counts to  the number of
$\phi\to\eta\gamma$ with $\eta\to\pi^0\pi^0\pi^0$ events selected in the
same analyzed sample. We get:
\begin{equation}
  Br(\phi\to\eta\pi^0\gamma)=(6.92\pm 0.10_{stat.}\pm 0.20_{syst.})\times 10^{-5}
\label{br1}
\end{equation}
where the first uncertainty is statistical and contains also the background
subtraction contribution, while the second one is systematic and is mainly
due to the error on $Br(\phi\to\eta\gamma)$, used in the normalization, and
to the selection cuts. \\
In fig.\ref{neutri}, right the distribution of the polar angle of
the recoil photon ({\it i.e.} the non associated one) after background
subtraction is also shown. The $1 + cos^2\Theta$ angular dependence, as expected for
the radiative decay of the $\phi$ into a scalar particle, is found.

\section{$\phi\to\eta\pi^0\gamma$ with $\eta\to\pi^+\pi^-\pi^0$}
Events from this decay chain are characterized by two tracks coming from the
interaction region and five
prompt photons. 
They are selected by requiring two tracks in the drift chamber coming from
a vertex close to the beam interaction point and five prompt photons, 
according to the definition given above.
There are no other relevant processes with exactly the same final state, the
background comes mainly from:
\begin{itemize}
  \item $e^+e^-\rightarrow\omega\pi^0$ with $\omega\rightarrow\pi^+\pi^-\pi^0$
    if one additional neutral cluster is given by the background or if one of
    the photons splits in two clusters;
  \item $\phi\rightarrow K_SK_L$ with prompt $K_L$ decay either in $3\pi^0$
    if $K_S\rightarrow\pi^+\pi^-$ (2 tracks and 6 photons), or in $\pi^+\pi^-\pi^0$, $\pi$l$\nu$ if
    $K_S\rightarrow\pi^0\pi^0$ (2 tracks and 6 or 4 photons): these two processes can mimic the signal 
    either if one photon is lost or if an additional accidental cluster is
    identified as a photon.
\end{itemize}
The background reduction is based on kinematic fitting using as constraints the
4-momentum conservation and the $\pi^0$ and $\eta$ invariant masses. The
overall efficiency for the signal is close to 20\%, while the residual
background is reduced down to approximately 15\% of the signal. The latter is
evaluated comparing data with MC for control samples dominated by the
different background components. 
At the end 4180 events are selected out of which we estimate a background contamination of 542 $\pm$ 57 events. 

\begin{figure}[htb]
\begin{center}
\mbox{\epsfig{file=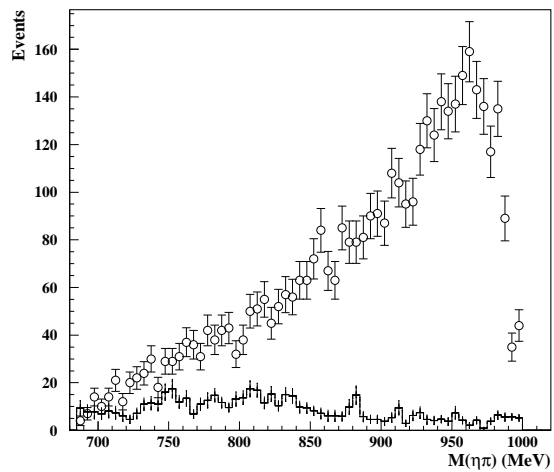,width=8cm}}
\end{center}
\caption{\small $\eta\pi^0$ raw mass spectrum of the selected events (open
  circles) compared to the estimated background (histogram) to be
  subtracted.}
\label{carichi}
\end{figure}

In fig.\ref{carichi} the $\eta\pi^0$ raw mass spectrum of the selected events is
compared to the same distribution for the estimated background. \\
After normalizing to $\phi\rightarrow\eta\gamma$ with 
$\eta\rightarrow\pi^0\pi^0\pi^0$ events (see
above) we get the branching ratio:    
\begin{equation}
  Br(\phi\to\eta\pi^0\gamma)=(7.19\pm 0.17_{stat.}\pm 0.24_{syst.})\times 10^{-5}
\label{br2}
\end{equation}
where the statistical error includes the background
subtraction while the the systematic error is mainly due to the
normalization procedure and to the efficiency evaluation. \\
Branching ratios (\ref{br1}) and (\ref{br2}) are independently measured
apart from the common normalization. 
In order to compare the two measurements, we recompute the error taking away systematics due to common sources. We obtain the values:
\begin{displaymath}
\begin{array}{lll}
  Br(\phi\to\eta\pi^0\gamma)=(6.92\pm 0.17)\times 10^{-5} & &
  (\eta\to\gamma\gamma) \\
  Br(\phi\to\eta\pi^0\gamma)=(7.19\pm 0.24)\times 10^{-5} & &
  (\eta\to\pi^+\pi^-\pi^0) 
\end{array}
\end{displaymath}
that are in good agreement.

\section{Fit of the invariant mass distribution}
In order to extract the relevant parameters of the $a_0$(980) from the
spectra of fig.\ref{neutri} and fig.\ref{carichi}, we exploited two different
models to parametrize the $\phi\to\eta\pi^0\gamma$ amplitude:
\begin{itemize}
  \item the ``Kaon loop'' (KL) model\cite{achasov} in which the coupling of the
    $\phi$ to the scalar mesons occurs through the formation of a charged
    kaon loop;
  \item the ``No structure'' (NS) model\cite{maiani}, in which a point-like
    coupling of the $\phi$ to $a_0\gamma$ is considered, and the scalar
    amplitude is parametrized as a Breit-Wigner, interfering with a
    polynomial background. 
\end{itemize}
\begin{figure}[htb]
\begin{tabular}{cc}
\hskip -.7cm
\mbox{\epsfig{file=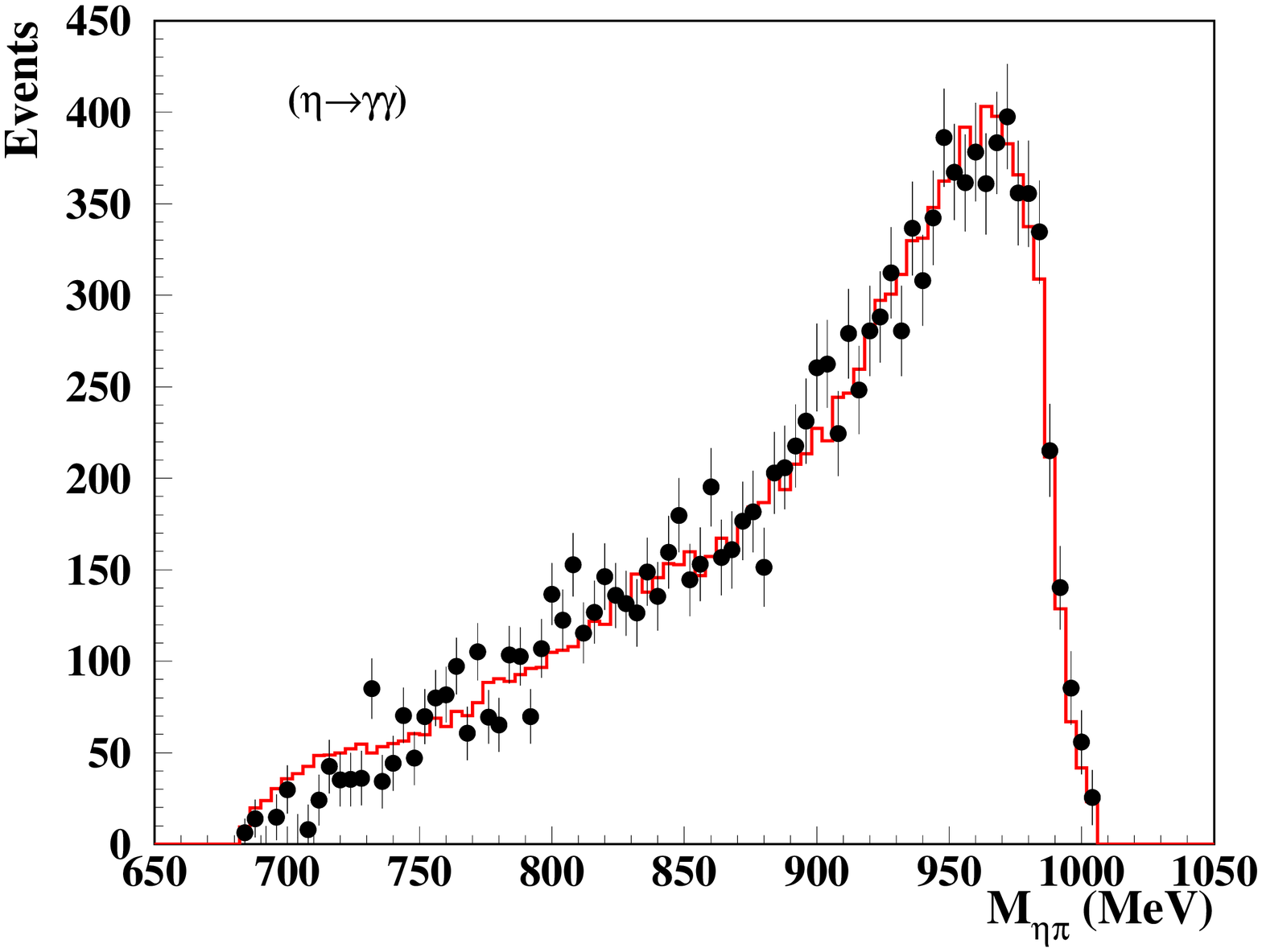, width=7cm}} &
\mbox{\epsfig{file=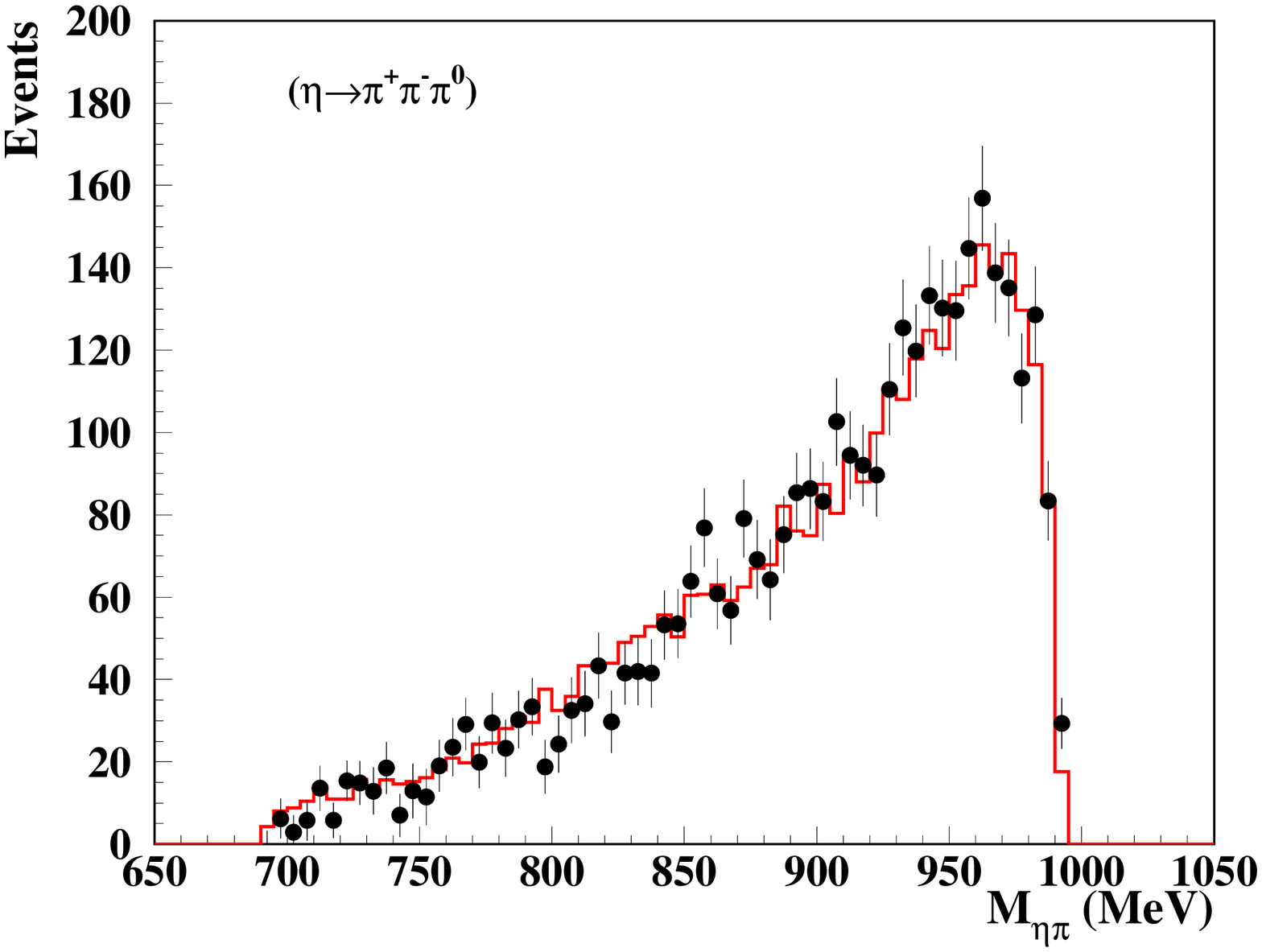, width=7cm}} \\
\end{tabular}
\caption{Combined fit for KL model. Black points: data; solid line: 
  fit result.}
\label{plotkl}
\end{figure}
A combined fit of the two spectra with the same parameters has been
performed.
In the KL case the free parameters are: the mass of the $a_0$, the couplings
$g_{a_0 K^+K^-}$ and $g_{a_0\eta\pi^0}$, the branching ratio of
$\phi\to\rho^0\pi^0\to\eta\pi^0\gamma$ and a phase $\delta$ between the scalar
and the vector amplitudes.
Another free parameter is the relative normalization between the two decay
channels, {\it i.e.} the ratio
$Br(\eta\to\gamma\gamma)/Br(\eta\to\pi^+\pi^-\pi^0)$.
The result of the fit is shown in fig.\ref{plotkl} while the parameters are
listed in tab.\ref{parameters}. \\
\begin{figure}[htb]
\begin{tabular}{cc}
\hskip -.7cm
\mbox{\epsfig{file=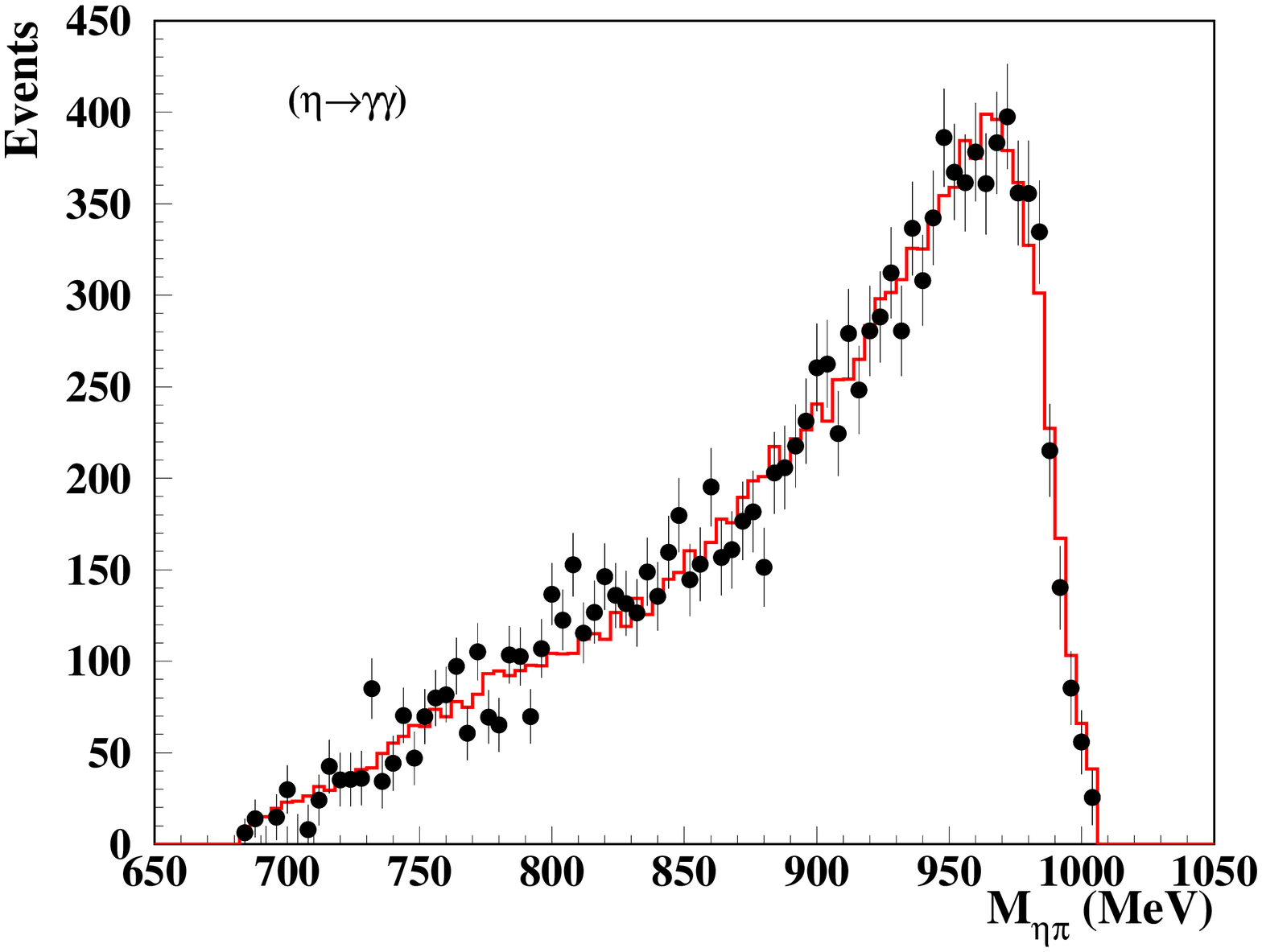, width=7cm}} &
\mbox{\epsfig{file=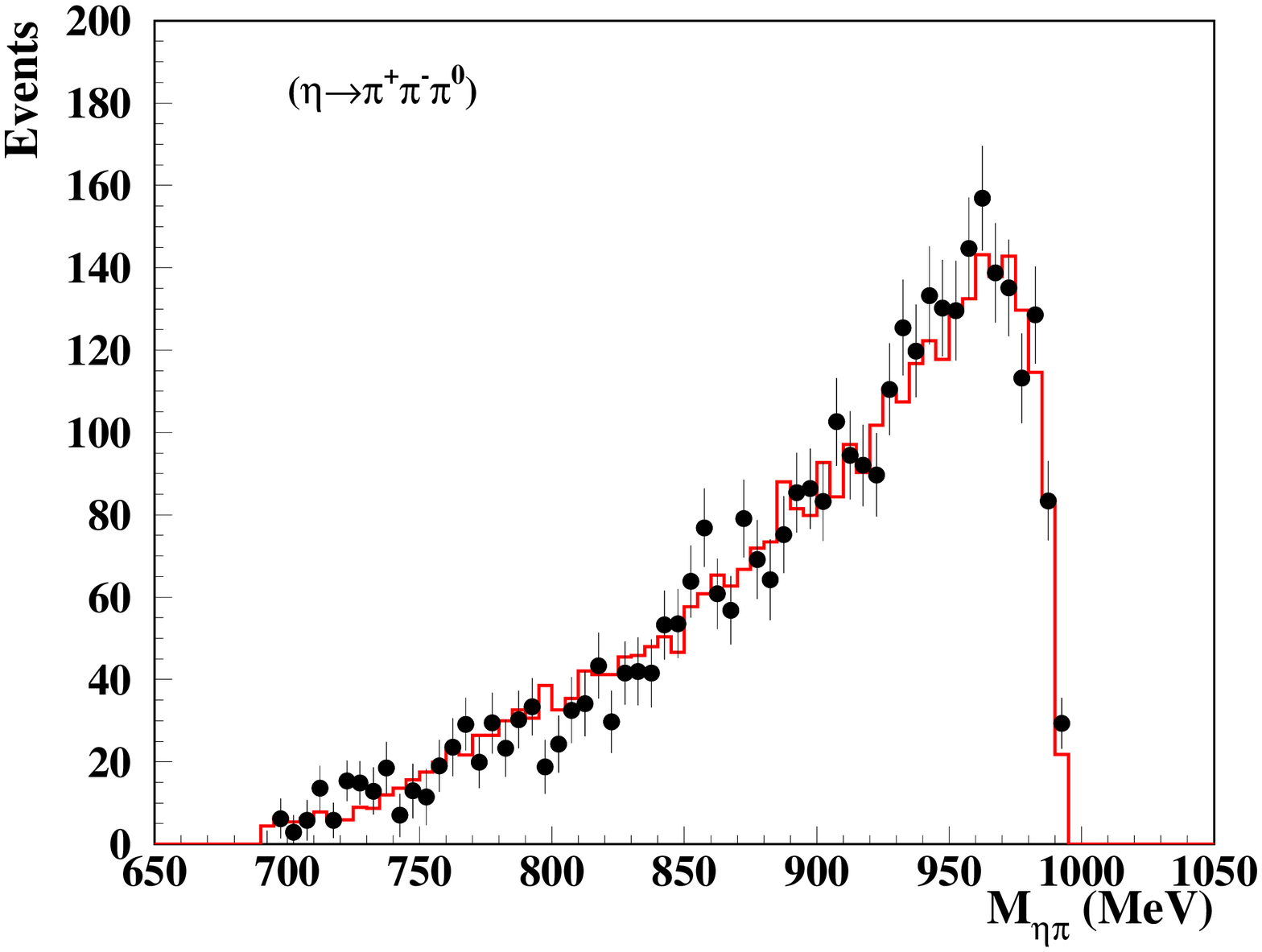, width=7cm}} \\
\end{tabular}
\caption{Combined fit for NS model. Black points: data; solid line: 
  fit result.}
\label{plotns}
\end{figure}
The number of free parameters of the NS model is larger: in addition to the same
parameters of the KL fit, there are also the $g_{\phi a_0\gamma}$ coupling  and 
two complex parameters for the polynomial background. In order to reduce
them, the $a_0$ mass has been fixed to 983 MeV, while the $\phi \to \rho \pi^0$
contribution has been set to the value calculated in
ref.\cite{maiani}.  
\begin{table}[htb]
\caption{Fit results (systematics on the parameters are not included).} 
\begin{tabular}{|l|c|c|}\hline
  & KL & NS \\
  \hline
  $M_{a_0}$ (MeV) & 983 $\pm$ 1 & 983 (fixed) \\
  $g_{a_0 K^+K^-}$ (GeV) & 2.16 $\pm$ 0.04 & 1.57 $\pm$ 0.13 \\
  $g_{a_0\eta\pi^0}$ (GeV) & 2.8 $\pm$ 0.1 & 2.2 $\pm$ 0.1 \\
  $g_{\phi a_0\gamma}$ (GeV$^{-1}$) & --- & 1.61 $\pm$ 0.05 \\
  $\delta (^{\circ})$ & 222 $\pm$ 12 & --- \\
  $Br(\phi\to\rho\pi^0\to\eta\pi^0\gamma)\times 10^6$ & 0.9 $\pm$ 0.4 &
  4.1 (fixed) \\
  $Br(\eta\to\gamma\gamma)/Br(\eta\to\pi^+\pi^-\pi^0)$ & 1.69 $\pm$ 0.04
  & 1.69 $\pm$ 0.04 \\
  $\chi^2$ & 156.6 & 146.8 \\
  $ndf$ & 136 & 134 \\
  $P(\chi^2,ndf)$ & 11\% & 21\% \\
  \hline
\end{tabular}
\label{parameters}
\end{table}
The results are plotted in fig.\ref{plotns} and the parameters values are
reported in tab.\ref{parameters}.
In both fits the relative normalization turns out to be in agreement, within the quoted errors,  with
the world average value (1.73$\pm$ 0.03 from PDG\cite{pdg}). 

\section{Conclusions}
In a sample of 414 pb$^{-1}$ of data we have measured the branching ratio of
$\phi\to\eta\pi^0\gamma$ selecting two different final states,
corresponding to $\eta\to\gamma\gamma$ and $\eta\to\pi^+\pi^-\pi^0$.
The two branching ratio values agree within the uncertainties. \\
From the combined fit of the $\eta\pi^0$ invariant mass distributions,
we obtain the values of the relevant parameters of the $a_0$(980).
Large value of $g_{a_0 K^+K^-}$ and $g_{\phi a_0\gamma}$ have been found,
suggesting a sizeable content of strange quark in the $a_0$(980).

\end{document}